\documentclass[aps,prd,onecolumn,groupedaddress,showpacs,nofootinbib,amssymb
]{revtex4}
\usepackage[dvips]{graphicx}
\usepackage{amssymb}
\usepackage{amsmath}
\usepackage{graphicx}
\usepackage{amsfonts}
\usepackage{bm}

\begin{document}

\title{Attractor Cosmology from non-minimally Coupled Gravity }
\author{S.D. Odintsov,$^{1,2,3,4}$\,\thanks{odintsov@ieec.uab.es}
V.K. Oikonomou,$^{5,6}$\,\thanks{v.k.oikonomou1979@gmail.com}}
\affiliation{$^{1)}$ ICREA, Passeig Luis Companys, 23, 08010 Barcelona, Spain\\
$^{2)}$ Institute of Space Sciences (ICE,CSIC) C. Can Magrans s/n,
08193 Barcelona, Spain\\
$^{3)}$ Institute of Space Sciences of Catalonia (IEEC),
Barcelona, Spain\\
$^{4)}$ Laboratory for Theoretical Cosmology, Tomsk State University
of Control Systems
and Radioelectronics, 634050 Tomsk, Russia (TUSUR)\\
$^{5)}$Department of Physics, Aristotle University of Thessaloniki, Thessaloniki 54124, Greece\\
$^{6)}$ Tomsk State Pedagogical University, 634061 Tomsk, Russia
}

\tolerance=5000

\begin{abstract}
By using a bottom-up reconstruction technique for non-minimally coupled scalar-tensor theories, we realize the Einstein frame attractor cosmologies in the $\Omega (\phi)$-Jordan frame. For our approach, what is needed for the reconstruction method to work is the functional form of the non-minimal coupling $\Omega(\phi)$ and of the scalar-to-tensor ratio, and also the assumption of the slow-roll inflation in the $\Omega (\phi)$-Jordan frame. By appropriately choosing the scalar-to-tensor ratio, we demonstrate that the observational indices of the attractor cosmologies can be realized directly in the $\Omega (\phi)$-Jordan frame. We investigate the special conditions that are required to hold true in for this realization to occur, and we provide the analytic form of the potential in the $\Omega (\phi)$-Jordan frame. Also, by performing a conformal transformation, we find the corresponding Einstein frame canonical scalar-tensor theory, and we calculate in detail the corresponding observational indices. The result indicates that although the spectral index of the primordial curvature perturbations is the same in the Jordan and Einstein frames, at leading order in the $e$-foldings number, the scalar-to-tensor ratio differs. We discuss the possible reasons behind this discrepancy, and we argue that the difference is due to some approximation we performed to the functional form of the potential in the Einstein frame, in order to obtain analytical results, and also due to the difference in the definition of the $e$-foldings number in the two frames, which is also pointed out in the related literature. Finally, we find the $F(R)$ gravity corresponding to the Einstein frame canonical scalar-tensor theory.
\end{abstract}

\pacs{04.50.Kd, 95.36.+x, 98.80.-k, 98.80.Cq,11.25.-w}

\maketitle



\def\pp{{\, \mid \hskip -1.5mm =}}
\def\cL{\mathcal{L}}
\def\be{\begin{equation}}
\def\ee{\end{equation}}
\def\bea{\begin{eqnarray}}
\def\eea{\end{eqnarray}}
\def\tr{\mathrm{tr}\, }
\def\nn{\nonumber \\}
\def\e{\mathrm{e}}

\section{Introduction}

Inflationary theories provide a successful description of the primordial Universe, since these solve in an appealing way the shortcomings of the Big Bang cosmology \cite{inflation2,inflation3,inflation4}. It was recently found that certain inflationary theories belong to a universality class, which provide a similar functional form of the observational indices in the Einstein frame. Examples of such theories are the $\alpha$-attractors \cite{alpha1} and similar forms, and they are extensively  studied in the literature, see \cite{alpha2,alpha3,alpha4,alpha5,alpha6,alpha7,alpha8,alpha9,alpha10,alpha11,alpha12,linderefs1,linder,Odintsov:2016vzz,Odintsov:2016jwr,extra1,extra2,extra3,extra4,extra5,extra6,extra7,extra8}. All these theories originate from different theories in the Jordan frame of a scalar field non-minimally coupled to gravity, to which we refer to hereafter as $\Omega(\phi)$-Jordan frame, but the appealing characteristic of these theories is that in the Einstein frame these yield inflation observational indices which are quite similar to each other, in the large-$N$ limit, where $N$ is the $e$-foldings number. Particularly, the spectral index of the primordial curvature perturbations $n_s$ and the scalar-to-tensor ratio have a simple form when these are calculated in the Einstein frame. Many well-known scalar-tensor inflationary theories belong to these attractor class of models, such as the Starobinsky model \cite{starob1,starob2} and the Higgs inflation model \cite{higgs}. A common characteristic of all the above models is that these yield a large plateau in their Einstein frame potential, which results to the similar behavior of the observational indices, in the large-$N$ limit, and for large field values.

In the spirit of attractor theories, in this paper we aim to study inflation in the $\Omega(\phi)$-Jordan frame, where a scalar field is non-minimally coupled to gravity. We shall assume that the slow-roll condition holds true directly in the $\Omega(\phi)$-Jordan frame, and by extending the bottom-up reconstruction method of Ref. \cite{frreconsbottomup}, for the case of non-minimally coupled $\Omega(\phi)R$ theories, we shall calculate in a general context the spectral index of the primordial curvature perturbations and the scalar-to-tensor ratio. As we will demonstrate, the resulting functional form of the observational indices in the $\Omega(\phi)R$ theory is quite similar to the cosmological attractors \cite{alpha2,alpha3,alpha4,alpha5,alpha6,alpha7,alpha8,alpha9,alpha10,alpha11,alpha12,linderefs1,linder,Odintsov:2016vzz,Odintsov:2016jwr,extra1,extra2,extra3,extra4,extra5,extra6,extra7,extra8}. We then transform the resulting $\Omega(\phi)R$ theory in the Einstein frame and we calculate the corresponding Einstein frame inflationary indices, by assuming there that the slow-roll approximation holds true. As we shall demonstrate, the resulting observational indices are not identical to the ones corresponding to the $\Omega(\phi)R$ theory. In principle, due to the fact that the spectral index and the scalar-to-tensor ratio are conformal invariant quantities, the resulting indices in the two frames should coincide \cite{kaizer,newsergei,extra3,Kuusk:2016rso}, so the different observational indices might be an artifact of the different slow-roll approximations in the two frames. It is worthy to note here that, differences occur frequently between the Jordan and Einstein frame when non-conformal invariant quantities are considered, for example the mass to radius ratio of neutron stars \cite{capp}, or acceleration in one frame may appear as deceleration in the other frame \cite{capp2}, in some cases finite-time singularities are of different types between the two frames \cite{noo5,bahamonte}, and also acceleration in the Jordan frame might appear as deceleration in the Einstein frame and vice-versa \cite{baha2}. In addition, the presence of matter can alter the resulting correspondence between conformal invariant quantities between the two frames, since in the Jordan frame matter is minimally coupled, while in the Einstein frame matter is not minimally coupled to the metric tensor. This issue is very delicate, so in a later section we shall concretely discuss this in a quantitative way.

Thus, by exploiting our bottom-up reconstruction method, in this work we shall demonstrate that the observational indices of the cosmological attractors can be obtained by using the $\Omega(\phi)$-Jordan frame directly, in the slow-roll approximation. Actually, we shall assume that the scalar-to-tensor has a specific desired functional form, and by using the reconstruction method we shall investigate to which $\Omega(\phi)$-Jordan frame theory this choice corresponds. Finally, we investigate which $F(R)$ gravity \cite{reviews1,reviews2,reviews3,reviews4,reviews5,reviews6} describes the resulting Einstein frame theory, thus connecting indirectly the $\Omega(\phi)$-Jordan frame with the $F(R)$ gravity frame theory.

This paper is organized as follows: In section II we present in brief the essential features of the cosmological attractors universality classes, and we discuss the similarities and overlaps between the various models and the similar Einstein frame picture. In section III we shall present in detail our bottom-up reconstruction technique for the $\Omega(\phi)$-Jordan frame theory, and we shall calculate the spectral index and the scalar-to-tensor ratio in the slow-roll approximation.  Moreover, we  examine the viability of the theory, and also the limitations of our approach. Finally, the conclusions follow at the end of the paper.

Before we start our presentation, let us briefly describe the geometric background which shall be assumed in the present paper and the conventions we shall assume to hold true. The geometric background will be assumed to be a flat Friedmann-Robertson-Walker (FRW), with the line element being,
\be
\label{metricfrw} ds^2 = - dt^2 + a(t)^2 \sum_{i=1,2,3}
\left(dx^i\right)^2\, , \ee
where $a(t)$ is the scale factor. Also the we assume that the connection is torsion-less, symmetric and metric compatible, the Levi-Civita connection. In addition, the Ricci scalar for the FRW metric (\ref{metricfrw}) is equal to,
\begin{equation}
\label{ricciscal} R=6(2H^2+\dot{H})\, ,
\end{equation}
where $H$ stands for the Hubble rate $H=\dot{a}/a$. Finally, we shall assume a physical units system such that $\hbar=c=8\pi G=\kappa^2=1$.

\section{Essential Features of Cosmological Attractor Models: $\Omega(\phi)$-Jordan Frame vs Einstein Frame}

The appealing properties of all the cosmological attractor models is that these yield a particular set of observational indices in the Einstein frame, which are,
\begin{equation}\label{einsteinframeattractorindices}
n_s=1-\frac{2}{N},\,\,\,r=\frac{12\alpha}{N^2}\, ,
\end{equation}
where the parameter $\alpha$ depends on the model chosen in the $\Omega(\phi)$-Jordan frame. Also as was demonstrated in \cite{alpha4}, the overlap of the cosmological models is due to the existence of a leading pole in the Laurent expansion of the kinetic term in the Einstein frame. So all the known attractor models, like the conformal attractors, $\xi$-attractors, $\alpha$-attractors and $\eta$-attractors and variants of these, have overlap even in their  $\Omega(\phi)$-Jordan frame corresponding theory, apart from the fact that in the Einstein frame, these lead to the same form of the observational indices (\ref{einsteinframeattractorindices}). Two interesting classes of attractor models are the $\xi$ attractor models and the $\eta$-attractor models, which originate from the following $\Omega(\phi)$-Jordan frame action,
\begin{equation}\label{omegafigeneralizedaction}
\mathcal{S}=\int \mathrm{d}^4x\sqrt{-g}\left(\Omega (\phi)R-\frac{K_J(\phi)}{2}\partial^{\mu}\phi\partial_{\mu}\phi-V(\phi)\right)\, ,
\end{equation}
where $K_J$ is the kinetic term in the $\Omega(\phi)$-Jordan frame. When this theory is conformally transformed to the Einstein frame, by applying the following transformation,
\begin{equation}\label{conformaltranf}
 \bar{g}_{\mu \nu}=\Omega g_{\mu\nu}\, ,
\end{equation}
where $\bar{g}_{\mu \nu}$ denotes the Einstein frame metric, we obtain the following Einstein frame action,
\begin{equation}\label{einsteinframelagrangian}
\mathcal{S}_E=\mathcal{S}=\int \mathrm{d}^4x\sqrt{-\bar{g}}\left(\bar{R}-\frac{\mathcal{K}_E(\phi)}{2}\partial^{\mu}\phi\partial_{\mu}\phi-\mathcal{V}_E(\phi)\right)\, ,
\end{equation}
where the Einstein frame kinetic term is related to the corresponding $\Omega(\phi)$-Jordan frame, as follows,
\begin{equation}\label{einsteinframekineticterm}
\mathcal{K}_E=\frac{K_J}{\Omega}+\frac{3}{2\Omega^2}\left( \frac{\mathrm{d}\Omega}{\mathrm{d}\phi}\right)^2\, ,
\end{equation}
and the Einstein frame potential is $\mathcal{V}_E=V/\Omega^2$. The $\xi$ attractor models and the $\eta$-attractor models correspond to different limiting cases of the $\Omega$ function, when $K_J=1$, with the $\xi$ attractor models being obtained in the limit $\Omega\gg 1$, with the corresponding Einstein frame potential being equal to,
\begin{equation}\label{einsteinframepotential}
\mathcal{V}_E(\phi)=V_0\left(1-\sum_{k=n}^{\infty}\xi_k\Omega^{-k} \right)\, ,
\end{equation}
where $\xi_k$ are constants and the leading order term in the above expansion is $\sim \Omega^{-n}$. The $\Omega(\phi)$-Jordan frame appearing in (\ref{einsteinframepotential}) can be made canonical if one makes the transformation $\chi=\sqrt{\frac{3}{2}}\ln \Omega$, so in the limit $\Omega\gg 1$, this would imply large and positive field values for the Einstein frame canonical scalar field $\chi$. In terms of the canonical scalar field, the leading order form of the Einstein frame potential is,
\begin{equation}\label{leadingeinsteinframepot}
\mathcal{V}_E(\chi)=V_0\left(1-\xi_n e^{-n\sqrt{\frac{2}{3}}\chi} \right)\, ,
\end{equation}
and the resulting Einstein frame observational indices, in the slow-roll approximation and at leading order, are equal to,
\begin{equation}\label{nsandreinsteinframexietaindices}
n_s=1-\frac{2}{N},\,\,\,r=\frac{12}{n^2N^2}\, .
\end{equation}
On the antipode of the $\xi$ attractor models, like the $\eta$-attractor models \cite{extra1}, in which case in the $\Omega(\phi)$-Jordan frame, the function $\Omega$ satisfies $\Omega\ll 1$, in which case the Einstein frame potential is,
\begin{equation}\label{einsteinframepotential2}
\mathcal{V}_E(\phi)=V_0\left(1-\sum_{k=n}^{\infty}\eta_k\Omega^{k} \right)\, ,
\end{equation}
and the leading term in this case is $\sim \Omega^{n}$, and $\eta_k$ are constants. If the scalar field $\phi$ is normalized, by using $\chi=\sqrt{\frac{3}{2}}\Omega$, the Einstein frame potential at leading order is the following,
\begin{equation}\label{leadingeinsteinframepot23}
\mathcal{V}_E(\chi)=V_0\left(1-\eta_n e^{n\sqrt{\frac{2}{3}}\chi} \right)\, ,
\end{equation}
but now since $\chi=\sqrt{\frac{3}{2}}\Omega$ and $\Omega\ll 1$, this implies $\chi$ is negative and takes very large values. Therefore, the resulting Einstein frame observational indices in this case too are given in Eq. (\ref{nsandreinsteinframexietaindices}), with the difference being that $n\neq 1$ in the $\xi$-attractor models.

Hence, having clarified the above essential issues of inflationary attractor models, one can understand the appealing property that these models have, which is a similar Einstein frame behavior. The aim of this paper is to produce the observational indices (\ref{nsandreinsteinframexietaindices}) directly from the $\Omega(\phi)$-Jordan frame theory, by assuming only that the slow-roll approximation holds true, and we also aim to compare the resulting theory with the corresponding Einstein frame one, in order to see how the observational indices behave in the two frames. As we already mentioned in the introduction, in principle these should be the same in the two frames, however the different slow-roll approximations may eventually lead to different results at leading $N$ order.

\section{$\Omega(\phi)R$ Gravity and Observational Indices of Attractor Models}

In this section we shall present in brief our bottom-up reconstruction method in the context of $\Omega(\phi)R$ gravity, which will enable us to generate the observational indices of cosmological attractors. The formalism is quite general, and in principle a large variety of functional forms of the observational indices can be generated. Let us briefly present the $\Omega(\phi)R$ gravity theoretical framework, and we specify the results in the context of the slow-roll approximation. The non-minimally coupled $\Omega(\phi)R$ gravity is described by the following action,
\begin{equation}
\label{graviactionnonminimal}
\mathcal{S}=\int d ^4x\sqrt{-g}\left(
\frac{\Omega(\phi)R}{2}-\frac{1}{2}g^{\mu\nu}
\partial_{\mu}\phi\partial_{\nu}\phi-V_J(\phi) \right)\, ,
\end{equation}
with $\Omega(\phi)$ being an analytic function of $\phi$. Upon variation of the gravitational action (\ref{graviactionnonminimal}) with respect to $\phi$, the following equations of motion are obtained,
\begin{equation}
\label{gravieqnsnonminimal}
3\Omega H^2=\frac{\dot{\phi}^2}{2}
+V_J(\phi)-3 H \dot{\Omega}\, ,\quad
 -2\Omega \dot{H}=\dot{\phi}^2
+\ddot{\Omega}-H \dot{\Omega}\, ,\quad
\ddot{\phi}+3H\dot{\phi}
 -\frac{1}{2 }R\frac{\mathrm{d}\Omega}{\mathrm{d} \phi}+\frac{\mathrm{d} V_J}{\mathrm{d} \phi}=0\, ,
\end{equation}
where the ``dot'' denotes differentiation with respect
to the cosmic time $t$. The dynamical evolution of inflation is perfectly described by the slow-roll indices, which for the non-minimally coupled theory have the following form \cite{reviews1},
\begin{equation}
\label{slowrollnonminimal}
\epsilon_1=-\frac{\dot{H}}{H^2}\, ,\quad
\epsilon_2=\frac{\ddot{\phi}}{H\dot{\phi}}\, ,\quad
\epsilon_3=\frac{\dot{\Omega}}{2H\Omega} \, ,\quad
\epsilon_4=\frac{\dot{E}}{2HE}\, ,
\end{equation}
where the function $E$ defined as follows,
\begin{equation}
\label{epsilonparameter}
E=\Omega+\frac{3\dot{\Omega}^2}{2 \dot{\phi}^2}\, .
\end{equation}
If the slow-roll indices satisfy $\epsilon_i\ll 1$,
$i=1,..,4$, in which case the slow-roll condition holds true, the observational indices of inflation can be expressed in terms of the slow-roll indices as follows,
\begin{equation}
\label{observatinalindices1}
n_s\simeq
1-4\epsilon_1-2\epsilon_2+2\epsilon_3-2\epsilon_4\, ,\quad
r=8 \frac{Q_s}{\Omega}\, ,
\end{equation}
where $n_s$ is the spectral index of the primordial curvature perturbations, $r$ is the scalar-to-tensor ratio and the function $Q_s$ is equal to,
\begin{equation}
\label{qs1nonslowroll}
Q_s=\dot{\phi}^2\frac{E}{\Omega H^2(1+\epsilon_3)^2}\, .
\end{equation}
If the slow-roll conditions are applied on the scalar field, the equations of motion take the following form,
\begin{align}
\label{gravieqnsslowrollapprx12}
& 3\Omega (\phi) H^2\simeq V_J(\phi)\, ,\quad
3H\dot{\phi}-6H^2\Omega'+V_J'\simeq 0\, , \\
\label{gravieqnsslowrollapprx3}
& \dot{\phi}^2\simeq
H\dot{\Omega}-2\Omega \dot{H}\, ,
\end{align}
where the ``prime'' hereafter denotes differentiation with respect
to the stated argument of the given function, for example above it denotes differentiation with respect to $\phi$. The above form of the equations of motion is particular useful for our reconstruction method, as we show shortly. By taking into account Eqs. (\ref{gravieqnsslowrollapprx3}), the function $Q_s$ can be written as follows,
\begin{equation}
\label{qs2}
Q_s=\frac{\dot{\phi}^2}{H^2}+\frac{3\dot{\Omega}^2}{2\Omega H^2}\, ,
\end{equation}
and by using Eq.~(\ref{gravieqnsslowrollapprx3}), we finally obtain,
\begin{equation}
\label{qs3}
Q_s\simeq
\frac{H\dot{\Omega}}{H^2}-\frac{2\Omega \dot{H}}{H^2}\, .
\end{equation}
In view of Eqs. (\ref{observatinalindices1}) and (\ref{qs3}), the scalar-to-tensor ratio reads,
\begin{equation}
\label{rscalartensornoniminal}
r\simeq 16 (\epsilon_1+\epsilon_3)\, .
\end{equation}
Also, the spectral index can be simplified in the following way,
\begin{equation}
\label{nsintersmofslowroll}
n_s\simeq 1-2\epsilon_1\left (\frac{3H\dot{\Omega}}{\dot{\phi}^2}+2
\right)-2\epsilon_2-6\epsilon_3\left(
\frac{H\dot{\Omega}}{\dot{\phi}^2}-1\right)\, .
\end{equation}
Now let us utilize the above equations, in order to introduce the bottom-up reconstruction method.  Firstly, we shall express all the above quoted physical quantities as functions of the $e$-foldings number $N$, and also we shall use the physical units system in which $\hbar=c=8\pi G=\kappa^2=1$, as we already mentioned in the introduction. By using the differentiation rules,
\begin{equation}\label{trick1}
\frac{\mathrm{d}}{\mathrm{d}t}=H\frac{\mathrm{d}}{\mathrm{d}N},\,\,\,\frac{\mathrm{d}^2}{\mathrm{d}t^2}=H^2\frac{\mathrm{d}^2}{\mathrm{d}N^2}+H\frac{\mathrm{d}H}{\mathrm{d}N}\frac{\mathrm{d}}{\mathrm{d}N}\, ,
\end{equation}
which connect the cosmic time and the $e$-foldings number $N$, the slow-roll indices as functions of $N$ read,
\begin{align}\label{slowrollindicesfrphin}
& \epsilon_1(N)=-\frac{H'(N)}{H(N)},\,\,\, \epsilon_2(N)=\frac{H'(N)^2+H(N)^2 \phi ''(N)}{H(N)^2 \phi '(N)},\,\,\,\epsilon_3(N)=\frac{\Omega'(N)}{2 \Omega(N)}\, ,
\end{align}
and in effect, the spectral index $n_s$ reads,
\begin{equation}\label{spcectrralindexnfrphi}
n_s=-2 \epsilon_1(N) \left(\frac{3 H(N)^2 \Omega'(N)}{H(N)^2 \phi '(N)^2}+2\right)-6 \epsilon_3(N) \left(\frac{H(N)^2 \Omega'(N)}{H(N)^2 \phi '(N)^2}-1\right)-2 \epsilon_2(N)+1\, ,
\end{equation}
and the scalar-to-tensor ratio is again given in Eq. (\ref{rscalartensornoniminal}).

The bottom-up reconstruction method we shall use in this work, is described in brief as follows: Firstly, the function $\Omega(N)$ is fixed to have a particular form, and also the scalar-tensor-ratio is also chosen to have a desirable form, for example $r=g(N)$. The choices of both $r=g(N)$ and $\Omega(N)$ that are made, can then be inserted in the expression for the scalar-to-tensor ratio (\ref{rscalartensornoniminal}), and the resulting differential equation will yield the functional form of $H(N)$. At this point, one should be sure that the resulting Hubble rate indeed describes an inflationary era, so one needs to solve the differential equation $\dot{N}=H(N(t))$, so the function $N(t)$ can be found. In effect, upon substitution of $N(t)$ in $H(N(t))$, the Hubble rate as a function of the cosmic time can be obtained, and then it can be easily checked if the resulting cosmological evolution satisfies $\ddot{a}>0$, in which case acceleration is achieved and an inflationary cosmology is obtained. Having $H(N)$ at hand, by solving the differential equation (\ref{gravieqnsslowrollapprx3}) one obtains $\phi(N)$, and if inverted, and substituting the resulting function $N^{-1}(\phi)$ in Eq. (\ref{gravieqnsslowrollapprx12}), the approximate form of the potential $V(\phi)$ can be found, during the slow-roll era. Also by substituting $H(N)$, $\Omega(N)$ and $\phi(N)$ in Eqs. (\ref{slowrollindicesfrphin}) and (\ref{spcectrralindexnfrphi}), we obtain the analytic form of $n_s$ as a function $N$.

  Let us provide in detail the formulas of the bottom-up reconstruction method, since these are useful for the rest of the paper. By combining Eqs. (\ref{slowrollindicesfrphin}) and (\ref{rscalartensornoniminal}), the following differential equation is obtained,
\begin{equation}\label{diffeqtionfrphiscalar}
16 \left(\frac{\Omega'(N)}{2 \Omega(N)}-\frac{H'(N)}{H(N)}\right)=g(N)\, .
\end{equation}
Given the non-minimal coupling function $\Omega(N)$, by solving the above, the analytic form of the Hubble rate as a function of $N$ is obtained. Then, $\phi(N)$ can be found, by solving the differential equation (\ref{gravieqnsslowrollapprx3}), which can be written as follows,
\begin{equation}\label{slowrollequatidifffrphi}
\left(H(N) \phi '(N)\right)^2=H(N)^2 \Omega'(N)-2 \Omega(N) H(N)H'(N)\, .
\end{equation}
By solving (\ref{slowrollequatidifffrphi}), the resulting function $\phi (N)$ can be substituted in the slow-roll index $\epsilon_2(N)$, and in effect, all the slow-roll indices can be expressed as functions of $N$, and hence the spectral index can be obtained as function of $N$.

Let us utilize the bottom-up reconstruction method we just presented, in order to realize the attractor cosmology observational indices of Eqs. (\ref{einsteinframeattractorindices}) and (\ref{nsandreinsteinframexietaindices}). As we shall demonstrate, there is no restriction on the parameters $n$ or $\alpha$ in our case, since these can be freely chosen in the context of the reconstruction formalism we shall use.

Let us assume that the $\Omega(\phi)$ function has the following form,
\begin{equation}\label{hubbleratemainassumption}
\Omega(N)=\beta  N^{\delta }\, ,
\end{equation}
where $\beta>0$ and the parameter $\delta$ will be specified later on. Also we assume that the scalar-to-tensor ratio has the following form,
\begin{equation}\label{scalartotensorratiofrphi}
r=\frac{c}{N^2}\, ,
\end{equation}
with $c>0$. The Eqs. (\ref{hubbleratemainassumption}) and (\ref{scalartotensorratiofrphi}) are the input equations of the bottom-up reconstruction method we presented earlier, and the rest of the quantities will be determined by using these equations. Indeed, by substituting Eqs. (\ref{hubbleratemainassumption}) and (\ref{scalartotensorratiofrphi}) in the differential equation (\ref{diffeqtionfrphiscalar}), with $g(N)=\frac{c}{N^2}$, by solving it, we obtain the Hubble rate, which is,
\begin{equation}\label{hubbleratemain}
H(N)=\gamma  e^{\frac{c}{16 N}} N^{\delta /2}\, ,
\end{equation}
where $\gamma$ is an integration constant which we assume it to be $\gamma>0$. Let us now demonstrate that the resulting cosmological evolution satisfies $\ddot{a}$, so it describes successfully an inflationary era. Indeed, by solving the differential equation $\dot{N}=H(N(t))$, we obtain,
\begin{equation}\label{ntefoldings}
N(t)=2^{\frac{2}{\delta -2}} ((2-\delta ) (\mathcal{C}_2+\gamma  t))^{-\frac{2}{\delta -2}}\, ,
\end{equation}
where $\mathcal{C}_2$ is an integration constant. Accordingly, the Hubble rate as a function of the cosmic time reads,
\begin{equation}\label{hubbleratecosmictime}
H(t)=\gamma  2^{\frac{\delta }{\delta -2}} (2-\delta )^{-\frac{\delta }{\delta -2}} (\mathcal{C}_2+\gamma  t)^{-\frac{\delta }{\delta -2}}\, .
\end{equation}
The scale factor can easily be found and it reads,
\begin{equation}\label{scalefactorfinished}
a(t)=\mathcal{C}_3 \exp \left(2^{\frac{4-\delta }{\delta -2}} (\delta -2)^{-\frac{2}{\delta -2}} (\mathcal{C}_2+\gamma  t)^{-\frac{2}{\delta -2}}\right)\, ,
\end{equation}
where $\mathcal{C}_3>0$ is an integration constant. So as we shall show, the scale factor satisfies $\ddot{a}$ for a wide range of the parameters, when the cosmic time is of the order $t\sim 10^{-36}-10^{-15}$sec, which is the cosmic time that inflation occurs in most grand unified theories.

By using Eqs. (\ref{hubbleratemainassumption}), (\ref{scalartotensorratiofrphi}) and (\ref{hubbleratemain}), and substituting in the differential equation (\ref{slowrollequatidifffrphi}), by solving the latter we obtain $\phi (N)$, which reads,
\begin{equation}\label{phinfrphi}
\phi (N)=\frac{\sqrt{\beta } \sqrt{c} N^{\delta /2}}{\sqrt{2} \delta }+\mathcal{C}_1
\, ,
\end{equation}
where $\mathcal{C}_1>0$ is an integration constant. Having the function $\phi(N)$ at hand, we can obtain the analytic form of all the slow-roll indices as functions of the $e$-foldings number, which read,
\begin{align}\label{slowrollindicesanalyticformfrphi}
& \epsilon_1(N)=\frac{c-8 \delta  N}{16 N^2},\,\,\,\epsilon_3(N)=\frac{\delta }{2 N},\\ \notag &\epsilon_2(N)=\frac{c^{3/2} N^{-\frac{\delta }{2}-3}}{64 \sqrt{2} \sqrt{\beta }}+\frac{\delta ^2 N^{-\frac{\delta }{2}-1}}{\sqrt{2} \sqrt{\beta } \sqrt{c}}-\frac{\sqrt{c} \delta  N^{-\frac{\delta }{2}-2}}{4 \sqrt{2} \sqrt{\beta }}+\frac{\delta }{2 N}-\frac{1}{N}\, .
\end{align}
The slow-roll indices $\epsilon_1(N)$ and $\epsilon_3(N)$ satisfy the slow-roll conditions in the large $N$ limit, if $N\gg \delta,\,\,c$, while the slow-roll index $\epsilon_2(N)$ when $\beta\gg N$ in conjunction with the conditions on $c$ and $\delta$. Hence, under these assumptions, the spectral index of Eq. (\ref{spcectrralindexnfrphi}) reads,
\begin{equation}\label{spectralindexanalyticformfrphi}
n_s=1-\frac{c}{4 N^2}+\frac{\delta }{N}+\frac{2}{N}\, ,
\end{equation}
while the scalar-to-tensor ratio is given in Eq. (\ref{scalartotensorratiofrphi}). Finally by choosing, $\delta=-4$, the spectral index reads,
\begin{equation}\label{nsspecialcase}
n_s\simeq 1-\frac{2}{N}\, ,
\end{equation}
while by choosing $c=\frac{12}{n^2}$ or $c=12\alpha$, the scalar-to-tensor ratio becomes identical to the $\eta$-attractors case of Eq. (\ref{nsandreinsteinframexietaindices}), or it becomes identical to the $\alpha$-attractors case of Eq. (\ref{einsteinframeattractorindices}), respectively. Note that in our case, the parameters $\alpha$ and $n$ can freely be chosen, in contrast to the Einstein frame theories of the $\eta$-attractors and $\alpha$-attractors.

At this point, we can calculate the scalar potential $V_J(\phi)$ of the $\Omega(\phi)$-Jordan frame theory in the slow-roll approximation, by combining Eqs. (\ref{gravieqnsslowrollapprx12}), (\ref{hubbleratemainassumption}) and (\ref{hubbleratemain}), and also by inverting Eq. (\ref{phinfrphi}), so the resulting expression in terms of the scalar field $\phi$ reads,
\begin{equation}\label{scalarjordanframepotential}
V_J(\phi)\simeq \frac{3 \gamma ^2 2^{1/\delta } \delta ^4 (\phi -\mathcal{C}_1)^4 \exp \left(2^{-\frac{1}{\delta }-3}c  \left(\frac{\delta  (\phi -\mathcal{C}_1)}{\sqrt{\beta } \sqrt{c}}\right)^{-2/\delta }\right)}{\beta  c^2}\, ,
\end{equation}
and accordingly, the function $\Omega (\phi)$ can be found in the same way and it reads,
\begin{equation}\label{omegaphifunction}
\Omega (\phi)=\frac{2^{1/\delta } \delta ^2 (\phi -\mathcal{C}_1)^2}{c}\, .
\end{equation}
Thus we demonstrated that if the $\Omega (\phi) $ theory satisfies certain constraints, it can lead to Jordan frame observational indices which can be identical to the ones corresponding to the $\eta$-attractors and $\alpha$-attractors. Also the bottom-up reconstruction method we used can result to a large variety of models, since the choice of $\Omega (\phi)$ can freely be made, and the same applies for the scalar-to-tensor ratio.
\begin{table*}[h]
\small \caption{\label{dosimo}The Quantitative Features of the Resulting $\Omega (\phi)$ Theory}
\begin{tabular}{@{}crrrrrrrrrrr@{}}
\tableline \tableline \tableline
 Scalar Potential $\quad$
 \\
$V_J(\varphi )\simeq  \frac{3 \gamma ^2 2^{1/\delta } \delta ^4 (\phi -\mathcal{C}_1)^4 \exp \left(c 2^{-\frac{1}{\delta }-3} \left(\frac{\delta  (\phi -\mathcal{C}_1)}{\sqrt{\beta } \sqrt{c}}\right)^{-2/\delta }\right)}{\beta  c^2}$
\\\tableline
$\Omega (\phi)$ Function
\\
$\Omega (\phi)=\frac{2^{1/\delta } \delta ^2 (\phi -\mathcal{C}_1)^2}{c}$
\\\tableline
Observational Indices in the $\Omega(\phi)$-Jordan frame$\quad $
\\
 When $\delta=-4$, $N\gg \delta $, $N\gg c$, $\beta\gg N$, $n_s\simeq 1-\frac{2}{N}$ and $r=\frac{c}{N^2}$
\\\tableline
\tableline \tableline
\end{tabular}
\end{table*}
In Table \ref{dosimo} we gather the details of the resulting $\Omega (\phi) $ theory, including the predicted observational indices. Before closing this section, we need to note that by choosing $\delta=-4$, the resulting evolution (\ref{scalefactorfinished}) is actually inflationary, but we skip the details for brevity.

\section{The Einstein and $F(R)$ Gravity Frame Description of the $\Omega(\phi)R$ Theory}

In this section we shall consider the Einstein and $F(R)$ gravity frame of the $\Omega(\phi)$ theory we found in the previous section. As we already discussed in the introduction, the spectral index and the scalar-to-tensor ratio should be the same in all frames, since these quantities are conformal invariant. However, due to the fact that the slow-roll conditions in the two frames are different, this might eventually break the conformal invariance and this could introduce differences in the observational indices when these are calculated in different frames. Also, the approximations made in the potential during the slow-roll era, are possibly the source of potential differences between different frames of gravity. As we will demonstrate, there are some differences between the $\Omega(\phi)$ theory and the corresponding Einstein frame theory, however both theories are viable.

In order to find the Einstein frame theory which corresponds to the Jordan frame $\Omega(\phi)$ theory, we make the conformal transformation,
\begin{equation}\label{conformaltransjordaneinstein}
\bar{g}_{\mu \nu}=\Omega (\phi)g_{\mu \nu}\, ,
\end{equation}
where $g_{\mu \nu}$ is the Jordan frame metric while $\bar{g}_{\mu \nu}$ is the Einstein frame metric. Under the Weyl transformation (\ref{conformaltransjordaneinstein}), the action of the scalar field in the Einstein frame reads,
\begin{equation}\label{actionofsclareinstein}
\mathcal{S}_E=\int d ^4x\sqrt{-\bar{g}}\left(
\frac{\bar{R}}{2}-\frac{1}{2}\mathcal{K}_E(\phi)\bar{g}^{\mu\nu}
\partial_{\mu}\phi\partial_{\nu}\phi-\mathcal{V}_E(\phi) \right)\, ,
\end{equation}
where the kinetic term $\mathcal{K}_E$ in the Einstein frame reads,
\begin{equation}\label{kinetictermeinsteinframe}
\mathcal{K}_E=\frac{3}{2} \left(\frac{\Omega '(\phi )}{\Omega (\phi )}\right)^2+\frac{1}{\Omega (\phi )}\, ,
\end{equation}
and the scalar potential $\mathcal{V}_E$ is related to the corresponding $\Omega(\phi)$-Jordan frame one, as follows,
\begin{equation}\label{scalarpoeinsyten}
\mathcal{V}_E=\frac{V_J(\phi)}{\Omega{\phi}^2}\, .
\end{equation}
For the $\Omega(\phi)$ appearing in Eq. (\ref{omegaphifunction}), the kinetic term reads,
\begin{equation}\label{kinetictermeinstein1}
\mathcal{K}_E=\frac{\frac{c 2^{-1/\delta }}{\delta ^2}+6}{(\mathcal{C}_1-\phi )^2}\, ,
\end{equation}
while the scalar potential reads,
\begin{equation}\label{scalarpotentialeinstien1}
\mathcal{V}_E=\frac{3 \gamma ^2 2^{-1/\delta } \exp \left(c 2^{-\frac{1}{\delta }-3} \left(\frac{\delta  (\phi -\mathcal{C}_1)}{\sqrt{\beta } \sqrt{c}}\right)^{-2/\delta }\right)}{\beta }\, .
\end{equation}
By using the following transformation, the scalar field $\phi$ can be transformed to the canonical scalar field $\chi$,
\begin{equation}\label{canonicalscalartransformation}
\chi=\int^{\phi} \sqrt{\mathcal{K}_E}\mathrm{d}\phi\, ,
\end{equation}
so the scalar field $\phi$ is related to the scalar field $\chi$ as follows,
\begin{equation}\label{scalarfieldchiphi}
\phi=e^{-\frac{\chi }{\sqrt{\frac{c 2^{-1/\delta }}{\delta ^2}+6}}}\, .
\end{equation}
Therefore, the new Einstein frame action in terms of the scalar field $\chi$ reads,
\begin{equation}\label{actionofsclareinstein}
\mathcal{S}_E=\int d ^4x\sqrt{-\bar{g}}\left(
\frac{\bar{R}}{2}-\frac{1}{2}\bar{g}^{\mu\nu}
\partial_{\mu}\chi\partial_{\nu}\chi-\mathcal{V}_E(\chi) \right)\, ,
\end{equation}
where the scalar potential in terms of $\chi$, is equal to,
\begin{equation}\label{scalarfieldpotentialofchi}
\mathcal{V}_E=\frac{3 \gamma ^2 2^{2-\frac{2}{\delta }} \exp \left(2^{-\frac{1}{\delta }-3} (-\delta ) c^{\frac{1}{\delta }+1} \mathcal{C}^{1/\delta } e^{-\frac{2 \chi }{(-\delta ) \sqrt{\frac{c 2^{-1/\delta }}{\delta ^2}+6}}}\right)}{\mathcal{C}_1}\, .
\end{equation}
It is hopeless to try to find the dynamics of inflation in the Einstein frame corresponding to the Einstein frame potential (\ref{scalarfieldpotentialofchi}), so due to the fact that the inflationary era takes place at large $\chi$ values, the potential can be approximated as follows,
\begin{equation}\label{potentialsimplifiedineinsteinframe}
\mathcal{V}_E \simeq V_0 \left(1-V_p e^{-\alpha  \chi }\right)\, ,
\end{equation}
where $\mathcal{V}_0$, $\mathcal{V}_p$ and $\alpha$ are defined as follows,
\begin{equation}\label{mathcalvovppaprameters}
\mathcal{V}_0=\frac{3 \gamma ^2 2^{2-\frac{2}{\delta }}}{\mathcal{C}_1},\,\,\,\mathcal{V}_p=2^{-\frac{1}{\delta }-3} (-\delta ) c^{\frac{1}{\delta }+1} \mathcal{C}_1^{1/\delta },\,\,\,\alpha=\frac{2}{(-\delta ) \sqrt{\frac{c 2^{-1/\delta }}{\delta ^2}+6}}\, .
\end{equation}
The slow-roll indices $\epsilon$ and $\eta$, for the canonical scalar field $\chi$ are defined as follows (recall that we use a physical units system in which $\kappa=1$),
\begin{equation}\label{slowrolldefinitions}
\epsilon=\frac{1}{2}\left( \frac{\mathcal{V}_E'(\chi)}{\mathcal{V}_E(\chi)}\right)^2,\, \,\,\eta= \frac{\mathcal{V}_E''(\chi)}{\mathcal{V}_E(\chi)}\, ,
\end{equation}
By using the potential (\ref{potentialsimplifiedineinsteinframe}), the slow-roll indices become,
\begin{equation}\label{slowrollanalyticexpressions}
\epsilon\simeq \frac{\alpha ^2 V_p^2 e^{-2 \alpha  \chi}}{2 \left(1-V_p e^{\alpha  (-\chi)}\right)^2},\,\,\,\eta \simeq -\frac{\alpha ^2 V_p e^{-\alpha  \chi }}{1-V_p e^{-\alpha  \chi }}\, .
\end{equation}
By using the condition $\epsilon(\chi_f)=1$ at the end of inflation, and also the fact that,
\begin{equation}\label{neasfunctionofslowrollindices}
N=\int_{\chi_f}^{\chi_i}\frac{\mathcal{V}_E(\chi)}{\mathcal{V}_E'(\chi)}\mathrm{d}\chi\, ,
\end{equation}
where $\chi_f$ and $\chi_i$ are the values of the scalar field $\chi$ at the end of inflation and at the horizon crossing instance, we can express the slow-roll indices as functions of the $e$-foldings number $N$. By doing so, and due to the fact that the spectral index of the primordial curvature perturbations $n_s$ and the scalar-to-tensor ratio for a canonical scalar field are equal to,
\begin{equation}\label{slowrollindicesasfunctoionsofn}
n_s\simeq 1-6 \epsilon+2\eta,\,\,\,r=16 \epsilon\, ,
\end{equation}
the resulting observational indices as function of the $e$-foldings number $N$, at leading order in the $e$-foldings number, are equal to,
\begin{equation}\label{finalobservationalindicescanonicalscalarchi}
n_s\simeq 1-\frac{2}{N},\,\,\,r\simeq \frac{8}{\alpha ^2 N^2}\, .
\end{equation}


By directly comparing the observational indices (\ref{finalobservationalindicescanonicalscalarchi}) with the ones appearing in Eqs. (\ref{nsspecialcase}) and (\ref{scalartotensorratiofrphi}), it can be seen that the scalar-to-tensor ratio is different. This issue occurs possibly due to the fact that the slow-roll conditions probably imply different large-N expansions in the Einstein and $\Omega(\phi)$ frames, with the difference being due to the fact that the Einstein frame slow-roll conditions do not imply the $\Omega(\phi)$ frame conditions \cite{Kuusk:2016rso}. If the calculations are performed numerically, these should coincide, however, the analytic calculations in the two frames, lead to different results. In addition, as was noted by Kaiser \cite{newsergei}, the difference might be traced to various ambiguities of the scalar field quantum fluctuations in the Jordan and Einstein frames. We believe that it has to be the combination of the slow-roll definition in the two frames, and of the scalar field quantum fluctuations ambiguities, that lead to different results in the two frames.

If we notice the spectral index in the two frames, we can see that the two expressions coincide. This was also noticed by Kaiser \cite{newsergei}, and the difference in the resulting scalar-to-tensor ratio must be due to the effect of the slow-roll condition and of the scalar field quantum fluctuations ambiguities in the tensor and scalar gravitational waves modes. Note also that it is known that even at a quantum level, the on-shell the theories in the Jordan and Einstein frame are equivalent, at least when a regular background is considered, see \cite{Nojiri:2000ja,Kamenshchik:2014waa,Ruf:2017xon}. Also, at the classical level, as was demonstrated in Refs.  \cite{extra3,Kuusk:2016rso}, the resulting observational indices are frame independent up to second order in the slow-roll parameters. For a relevant approach on this, see also \cite{Jarv:2016sow}. Nevertheless, the results of \cite{Kuusk:2016rso} are obtained at the level of the slow-roll expansion, without invoking the $e$-foldings number. To our opinion, the solution to this theoretical problem is nicely presented in Ref. \cite{extra3}, in which the authors find a difference in the definition of the $e$-foldings number in the two frames, which in our case and by using our notation results to the following formula,
\begin{equation}\label{enisntejornaefoldingsconnection}
\mathrm{d}\bar{N}=\mathrm{d}N+\frac{1}{2}\mathrm{d}\ln(\frac{1}{\Omega(\chi)})\, ,
\end{equation}
where $\bar{N}$and $N$ are the Jordan frame and Einstein frame $e$-foldings numbers respectively. It is worth discussing further the discrepancies between different frames of gravity. The difference of the behavior of $H(N)$ and $N$ in different conformal frames of gravity, was pointed out some time ago in \cite{levin} in the context of kinetic inflation theories. Furthermore, the $\alpha$-attractor class of models seem to be specific examples of a quite general class of non-minimal coupling attractor models which were firstly studied in \cite{kaiserrevision}. Moreover, in Ref. \cite{tsuji}, the authors  found agreement for both $n_s$ and
$r$, which is different from the findings of our work, since there is some difference in the scalar-to-tensor ratio. As we discussed above, this might be due to the fact that there is difference between the definition of the $e$-foldings number in the two frames (and hence in expansions in terms of the $e$-folding number, of various physical quantities), and also due to the approximations we did in the Einstein frame in order to obtain an analytic form of the scalar potential.

As we demonstrated, in this work, it is possible to rescale the scalar field to have canonical kinetic terms in each
frame. If one works with canonically normalized scalar fields in each
frame, perhaps it is not surprising that quantities related to the
scalar perturbations remain frame-invariant, such as the spectral index in our case $n_s$. However,
the tensor perturbations in the Jordan frame involve significant
kinetic mixing between $h_{\mu\nu}$ and $\phi$. If one does not first diagonalize the kinetic sector for the tensor perturbations, before calculating quantities like the tensor-mode power spectrum (or
related ratios like the scalar-to-tensor ratio $r$), perhaps one should indeed find some residual frame
dependence. The importance of diagonalizing the
kinetic sector of the tensor modes, and not just the scalar field, was firstly pointed out clearly in Ref. \cite{berz}. Let us finally note that it is only in
single-field models with non-minimal couplings, that one may always
rescale the scalar field to make its kinetic energy canonical in
either frame. With two or more non-minimally coupled fields, such
rescalings are not possible anymore, and some non-canonical kinetic terms
necessarily remain in the Einstein frame, even if the scalar fields
kinetic sector is canonical in the Jordan frame, as was demonstrated in Ref. \cite{kaizerf}. Resolving the frame-dependence issue in the context of multi-field models is a challenge, since in these models even the power spectrum can show non-trivial frame-dependence, see for example \cite{sasaki}.

In a future work we shall further study the insightful result of \cite{extra3}, and further support the claim of the authors for the $e$-foldings dependence of the result. This will be done by using the general formulas of the slow-roll indices as functions of the $e$-foldings number, which we provided in the previous section. Thus in our case, the difference possibly comes from the definition of the $e$-foldings number in the two frames, but also possibly due to the approximation (\ref{potentialsimplifiedineinsteinframe}) which we assumed for the derivation of the resulting expression for the Einstein frame potential. In addition, in the case at hand, the difference between the two frames is traced only on the scalar-to-tensor ratio, but still, in both cases, the scalar-to-tensor ratio scales as $r\sim N^{-2}$, and the ratio $s=r_J/r_E$ of the predicted Jordan to Einstein frame scalar-to-tensor ratio is,
\begin{equation}\label{s}
s=2 c \sqrt{\frac{c}{8\ 2^{3/4}}+6}\, ,
\end{equation}
so if $c$ is small, the ratio $s$ is small.

\begin{figure}[h]
\centering
\includegraphics[width=22pc]{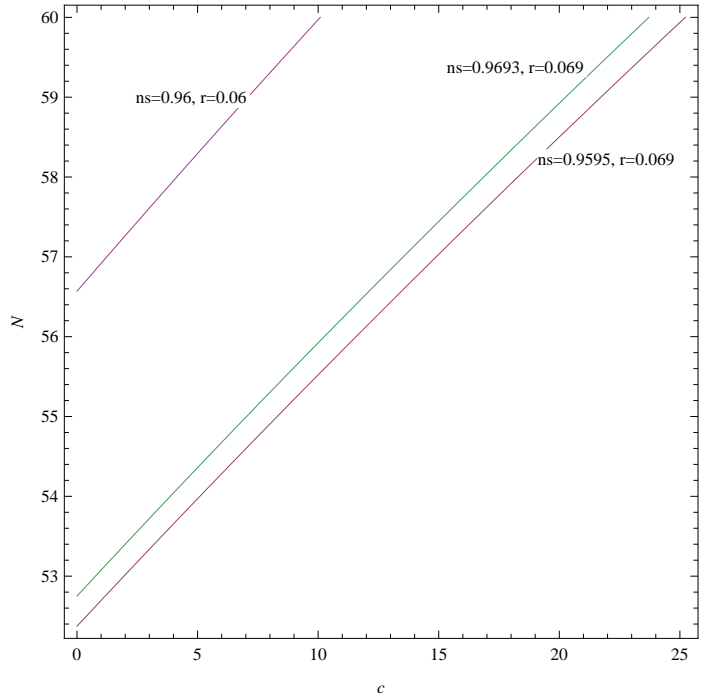}
\caption{The contour-plot of $n_s$ and $r$, as functions of $c$ and $N$, by using some characteristic values of $n_s$ and $r$. The upper curve corresponds to $n_s=0.96$ and $r=0.06$, the lower curve to $n_s=0.9595$ and $r=0.069$ and the middle curve to $n_s=0.9693$ and $r=0.069$.}\label{plot1}
\end{figure}

We can easily check the viability of the Einstein frame resulting theory by confronting the resulting observational indices with the latest Planck \cite{Ade:2015lrj} and BICEP2/Keck-Array data \cite{Array:2015xqh}. The 2015 Planck data \cite{Ade:2015lrj} impose the following constraints on $n_s$ and $r$,
\begin{equation}
\label{planckdata} n_s=0.9644\pm 0.0049\, , \quad r<0.10\, ,
\end{equation}
while the BICEP2/Keck-Array data \cite{Array:2015xqh} further constrain $r$ as follows,
\begin{equation}
\label{scalartotensorbicep2} r<0.07\, ,
\end{equation}
at $95\%$ confidence level. By choosing for example $N=60$ and $c=0.01$, the observational indices of Eq. (\ref{finalobservationalindicescanonicalscalarchi}) take the following values $n_s=0.96$ and $r=0.0533399$, which are compatible with both the Planck and BICEP2/Keck-Array data. In Fig. \ref{plot1}, we present the contour-plot of $n_s$ and $r$, as functions of $c$ and $N$, by using some characteristic values of $n_s$ and $r$. For the spectral index, the allowed range of values by the Planck data are in the interval $n_s=[0.9595,0.9693]$, and for $r<0.07$. The upper curve corresponds to $n_s=0.96$ and $r=0.06$, the lower curve to $n_s=0.9595$ and $r=0.069$ and the middle curve to $n_s=0.9693$ and $r=0.069$. As it can be seen, there is a wide range of the parameters $(c,N)$, for which the compatibility with the observational data is achieved.


As a final task, we shall find the $F(R)$ gravity theory which corresponds to the Einstein frame canonical scalar theory with the potential (\ref{potentialsimplifiedineinsteinframe}). The procedure is well-known, see for example the reviews \cite{reviews1,reviews2,reviews3,reviews4,reviews5,reviews6}. We start from the $F(R)$ gravity action,
\begin{equation}\label{pure}
\mathcal{S}=\frac{1}{2}\int\mathrm{d}^4x \sqrt{-\hat{g}}F(R)\, ,
\end{equation}
where $\hat{g}_{\mu \nu}$, is the $F(R)$ gravity frame metric tensor. By introducing the auxiliary scalar field $A$, the action (\ref{pure}) is written as follows,
\begin{equation}\label{action1dse111}
\mathcal{S}=\frac{1}{2}\int \mathrm{d}^4x\sqrt{-\hat{g}}\left (
F'(A)(R-A)+F(A) \right )\, .
\end{equation}
Upon variation of the action (\ref{action1dse111}) with respect to $A$, we obtain $A=\hat{R}$, so this solution explicitly proves that the gravitational actions (\ref{action1dse111})
and (\ref{pure}) are mathematically equivalent. The $F(R)$ gravity theory can be transformed in the Einstein frame one by making the following transformation,
\begin{equation}\label{can}
\chi =\sqrt{\frac{3}{2}}\ln (F'(A))
\end{equation}
with $\chi$ is the canonical scalar field of the Einstein
frame theory we found previously. The metric in the $F(\hat{R})$ frame can be transformed as follows,
\begin{equation}\label{conftransmetr}
\bar{g}_{\mu \nu}=e^{-\chi }\hat{g}_{\mu \nu }\, ,
\end{equation}
and the gravitational action under the transformation (\ref{conftransmetr}) becomes,
\begin{align}\label{einsteinframeaction}
& \mathcal{\tilde{S}}=\int \mathrm{d}^4x\sqrt{-\hat{g}}\left (
\hat{R}-\frac{1}{2}\left (\frac{F''(A)}{F'(A)}\right )^2g^{\mu \nu
}\partial_{\mu }A\partial_{\nu }A -\left (
\frac{A}{F'(A)}-\frac{F(A)}{F'(A)^2}\right ) \right ) \\ \notag & =
\int \mathrm{d}^4x\sqrt{-\hat{g}}\left ( \hat{R}-\frac{1}{2}g^{\mu \nu
}\partial_{\mu }\chi \partial_{\nu }\chi -V(\chi )\right )\, .
\end{align}
The Einstein frame scalar potential $V(\chi )$ is linked to the $F(\hat{R})$ gravity theory as follows,
\begin{align}\label{potentialvsigma}
V(\chi
)=\frac{1}{2}\left(\frac{A}{F'(A)}-\frac{F(A)}{F'(A)^2}\right)=\frac{1}{2}\left
( e^{-\sqrt{2/3}\chi }\hat{R}\left (e^{\sqrt{2/3}\chi} \right )-
e^{-2\sqrt{2/3}\chi }F\left [ \hat{R}\left (e^{\sqrt{2/3}\chi}
\right ) \right ]\right )\, .
\end{align}
The above relation is quite useful, since when the potential $V(\chi)$ is given, by combining Eqs. (\ref{potentialvsigma}) and (\ref{can}), we obtain the following differential equation,
\begin{equation}\label{solvequation}
\hat{R}F_{\hat{R}}=2\sqrt{\frac{3}{2}}\frac{\mathrm{d}}{\mathrm{d}\chi}\left(\frac{V(\chi)}{e^{-2\left(\sqrt{2/3}\right)\chi}}\right)
\end{equation}
with $F_{\hat{R}}=\frac{\mathrm{d}F(\hat{R})}{\mathrm{d}\hat{R}}$. In conjunction with Eq. (\ref{can}), the above equation will enable us to find the $F(\hat{R})$ gravity theory which corresponds to a given Einstein frame scalar potential $V(\chi)$. Let us now find the $F(\hat{R})$ gravity theory which corresponds to the potential (\ref{potentialsimplifiedineinsteinframe}), so by using $F_{\hat{R}}=e^{\sqrt{\frac{2}{3}}\chi}$, which can be obtained from Eq. (\ref{can}), we obtain the following algebraic equation,
\begin{equation}\label{staroalgebreqn}
F_{\hat{R}} \hat{R}-\sqrt{\frac{2}{3}} F_{\hat{R}}^2 V_0 F_{\hat{R}}^{\sqrt{\frac{3}{2}} (-\alpha )} \left(2 \sqrt{6} F_{\hat{R}}^{\sqrt{\frac{3}{2}} \alpha }+3 \alpha  V_p-2 \sqrt{6} V_p\right)=0\, .
\end{equation}
Due to the fact that $\chi\gg 1$ during the inflationary era, from Eq. (\ref{can}) we get that $F_{\hat{R}}\gg 1$ during inflation, so the algebraic equation (\ref{staroalgebreqn}) becomes,
\begin{equation}\label{algrebraicnew1}
F_{\hat{R}} \hat{R}-\sqrt{\frac{2}{3}} F_{\hat{R}}^2 V_0 F_{\hat{R}}^{\sqrt{\frac{3}{2}} (-\alpha )} \left(2 \sqrt{6} F_{\hat{R}}^{\sqrt{\frac{3}{2}} \alpha }\right)=0\, ,
\end{equation}
which can be solved and it yields,
\begin{equation}\label{canonicalfrjordanframe}
F_{\hat{R}}\simeq \frac{\hat{R}}{4 V_0}\, ,
\end{equation}
so by integrating with respect to the Ricci scalar, we obtain the approximate form of the $F(\hat{R})$ gravity during inflation, which is,
\begin{equation}\label{canonicalfrjordanframe1}
F(\hat{R})\simeq \frac{\hat{R}^2}{8 V_0}+\Lambda\, ,
\end{equation}
where $\Lambda$ is an integration constant. The result is similar to the one obtained in Ref. \cite{Odintsov:2016vzz}, and the similarity is due to the fact that the models originate from a similar exponential potential in the Einstein frame.

\section{Conclusions}

In this paper we studied how the $\alpha$ and $\eta$ attractors predictions can be realized by $\Omega (\phi)$ non-minimally coupled gravity, in the context of slow-roll approximation in the Jordan frame. In order to achieve this, we used the formalism of a bottom-up reconstruction technique, in which only the functional form of the scalar-to-tensor ratio and of the non-minimal coupling are considered known, and the rest of the quantities of the theory are derived by these two. We also provided analytic formulas of the slow-roll indices and of all the quantities involved in the theory, in terms of the $e$-foldings number. As we demonstrated, it is possible to realize directly from the $\Omega (\phi)$-Jordan frame the attractor cosmologies, and this is the novel feature of this work. In most of the attractor cosmologies, different $\Omega (\phi)$-Jordan frame theories lead to a similar potential in the Einstein frame, and the attractor property of having similar observational indices, can be obtained in the Einstein frame counterpart theory, while in our case the attractor observational indices can be obtained from the Jordan frame theory directly, with the only assumption made being that the slow-roll condition holds true. This result raises the question, why should the Jordan frame theory be studied, since due to the conformal invariance of the observational indices, these should be the same in both the Einstein and Jordan frame. In view of this, we performed a conformal transformation in the Einstein frame, and we calculated the resulting potential and the corresponding observational indices. As we demonstrated, the spectral index of the primordial curvature perturbations is indeed the same in the two frames, however, the scalar-to-tensor ratio is different. We extensively discussed the source of this difference, since in the literature there appear various contradicting proposals, and we concluded that the difference is due to a simplification of the potential which we assumed in the Einstein frame, in order to obtain analytic results, but also due to the fact that the definition of the $e$-foldings number in the two frames is different in the two frames. The latter is also bibliographically supported. In a future work we aim to address this issue with a focused work on that. In conclusion, the best way to examine the viability of an inflationary theory, is by directly calculating the observational indices, and this procedure will reveal any differences between the Jordan and Einstein frame counterpart scalar-tensor theories. In this way, with this work we showed that the predictions of inflation in the $\Omega (\phi)$-Jordan frame and the corresponding Einstein frame, are the same at leading order in the $e$-foldings number, when the spectral index is considered, however the predictions  on the scalar-to-tensor ratio differ.

\section*{Acknowledgments}

This work is supported by MINECO (Spain), FIS2016-76363-P (S.D.O) and by SGR247 (AGAUR, Catalonia), 2017 (S.D.O).

\end{document}